\documentclass[aps,prl,groupedaddress,twocolumn,showpacs]{revtex4}
\usepackage{amsmath}
\usepackage{amssymb}
\usepackage[final]{graphicx}
\begin{document}

%\preprint{}

\title{Optimal Synchronization in Space}

\author{Markus Brede}
%\email[]{Your e-mail addres}
%\thanks{}
\affiliation{CSIRO Marine and Atmospheric Research, CSIRO Centre for Complex System Science, F C Pye Laboratory,
GPO Box 3023, Clunies Ross Street
Canberra ACT 2601, Australia}

\email{Markus.Brede@Csiro.au}


\date{\today}

\begin{abstract}
In this Rapid Communication we investigate spatially constrained networks that realize optimal synchronization properties. After arguing that spatial constraints can be imposed by limiting the amount of `wire' available to connect nodes distributed in space, we use numerical optimization methods to construct networks that realize different trade-offs between optimal synchronization and spatial constraints. Over a large range of parameters such optimal networks are found to have a link length distribution characterized by power law tails $P(l)\propto l^{-\alpha}$, with exponents $\alpha$ increasing as the networks become more constrained in space. It is also shown that the optimal networks, which constitute a particular type of small world network, are characterized by the presence of nodes of distinctly larger than average degree around which long distance links are centred. 
\end{abstract}


\maketitle

%\section{Introduction}
In recent years much insight has been gained about the structure and function of many complex systems by mapping them to networks, nodes being associated with elementary units and links describing the coupling between them. Surprisingly, universal categories of networks have been found: scale-free \cite{Barabasi} and small world (SW) networks \cite{Strogatz} being the most important hallmarks. Recent review articles can be found in \cite{nets}. After considerable insight has been gained about the structure of many real-world networks, one important strand in network science is to understand how the dynamics of a system can be constrained by the structure of the coupling topology. Due to their wide range of applications including biological and ecological problem settings, opinion formation in social science, engineering applications and laser physics, synchronization problems on networks have recently attracted much attention in this field, cf. \cite{synrev} for a recent review. Note that spatial constraints play a role in almost all of these applications. 

Starting with the study of Watts and Strogatz \cite{Strogatz}, investigating the dynamics of synchronization of coupled oscillators has faciliated the discovery of important classes of networks, SW networks being the prime example. SW networks combine properties of the underlying spatial organization with those of random graphs and can be seen as arising from a tendency to minimize the cost to connect nodes distributed in space while at the same time trying to achieve shortest pathlengths on the network \cite{Gopal}. 

SWs have been associated with enhanced synchronization properties, but have later been shown not to be optimal for synchronization \cite{Donetti}. Reminiscent of the underlying spatial organisation, SWs are characterised by the presence of many short loops. In contrast, networks with optimal synchronisability are so-called entangled networks \cite{Donetti}. These networks, are regular and small, but have large girths, i.e. these networks are marked by the avoidance of small loops and thus also by the lack of local connections. By addressing the question: ``How many long links does a SW need to synchronize?'' we will investigate this apparent trade-off between spatial embeddedness and synchronization in this Rapid Communication. Tuning the constraining effect of space on network organization, we construct ensembles of synchrony-optimized networks, which combine properties of the underlying space and synchrony-optimality in different ways. We find that such networks are characterized by power laws in the link size distribution, thus replicating an interesting pattern that has, for instance, been observed in the human brain \cite{Schuz,Egui}. We also show that SWs can be networks with optimal synchronization properties, if the network organization is constrained by the underlying spatial structure.

Let us consider the dynamics given by the following equation, coupling the dynamics of oscillators each described by its individual dynamics given by $\dot{\phi_i}=f(\phi)$,
\begin{align}
\label{E0}
 \dot{\phi_i}=f(\phi_i)+\sigma \sum_j A_{ij} [g(\phi_j)-g(\phi_i)],
\end{align}
where the function $g$ describes the coupling, $\sigma$ the coupling strength and the matrix $A=(A_{ij})$ the adjacency matrix of the coupling network. In the following we consider connected undirected graphs corresponding to symmetrical adjacency matrices A. 

In an analysis of the stability of the fully synchronized state $f(s)=0$ of Eq. (\ref{E0}) Pecora and Carroll derived a `master stability' function, relating the stability of synchronized solutions to the eigenvalues of the graph Laplacian $L_{ij}=k_i \delta_{ij}-A_{ij}$, where $k_i=\sum_{j}A_{ij}$ \cite{Pecora}. Since we consider connected undirected graphs, the eigenvalues of $L$ are all real and there is exactly one zero eigenvalue. Let the eigenvalues of $L$ be labelled in ascending order $0=\lambda_0<\lambda_1\leq ...\leq \lambda_N$. According to \cite{Pecora} for many classes of oscillators the stability of the fully synchronized state is related to a small eigenratio $e=\lambda_N/\lambda_1$, such that a tightly packed spectrum of $L$ begets stability of the synchronized state. It needs to be emphasised that this eigenvalue ratio is only a measure for the stability of the synchronized state and does not a priori give any information about the onset or transition towards synchronization. However, its generality and independence of the particulars of the analyzed system has led to it being widely adopted as a measure for the `synchronizability' of a network, cf. \cite{synrev}. This notion is further supported by \cite{Chavez}, which demonstrates for a class of networks that a smaller eigenratio also implies an earlier onset of synchronization, even for non-identical units. For these reasons we follow previous studies and adopt the eigenratio as a measure for the synchronizability of networks. 

We consider a set of $N$ identical oscillators coupled by Eq. (\ref{E0}) that are uniformly distributed in a 1-dimensional space with periodic boundary conditions, e.g. with spatial locations $x_i=i \delta$, $i=0,...,N-1$. The parameter $\delta$ gives the distance between adjacent nodes and we fix minimum length scales to $\delta=1$ in the following considerations. In one dimension the distance between nodes $i$ and $j$ is then given by $d(i,j)=\min(|x_i-x_j|, N\delta-|x_i-x_j|)$.  Thus, the amount of `wire' required to realize the connections of a network topology $A$ is obtained from $W=\sum_{i<j} A_{ij} d(i,j)$.

Spatial constraints on the network organisation can be formulated as constraints on the amount of wire available to connect the network. The minimum amount of wire to connect all units is $(N-1)\delta$ corresponding to a linear chain that connects spatial nearest neighbours, while $\delta N(N^2-1)/16$ for a fully connected network gives the maximum amount of wire that can be used. If establishing physical connections is very expensive, i.e. only a very limited amount of wire is available, links will mostly be restricted to short, i.e. local connections. For expensive wire there is a pressure for networks to be organized as a spatial grid. On the other hand, if wire is cheap and plenty is available, no constraint towards local connections is enforced and the effects of the underlying spatial organization will not be important. Thus tuning the amount of available wire allows to systematically change the influence of spatial constraints on network organization.

\begin{figure*}
\begin{center}
\includegraphics [width=.9\textwidth]{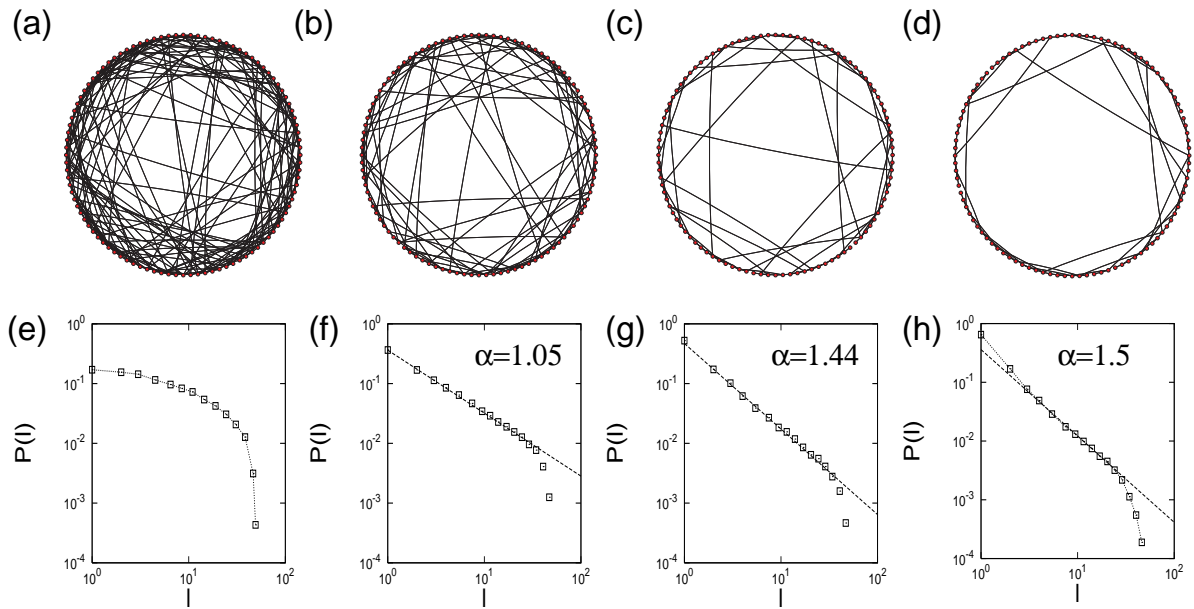}
\caption{(a)-(d): Example networks constructed for $\lambda=0.05$, $0.3$, $0.8$ and $0.95$. (e)-(h) Link size distributions for $\lambda=0.05$, $0.3$, $0.8$ and $0.95$. $N=100$, for each value of $\lambda$ an ensemble of $100$ networks has been constructed. In (e) and (h) points are connected to guide the eyes, straight lines in (f), (g) and (h) compare with power laws with exponents $\alpha=1.05$, $\alpha=1.44$ and $\alpha=1.5$, respectively. In panels (c) and (d) it can easily be observed that a few nodes attract a disproportionately large share of long links.}
\label{F3}
\end{center}
\end{figure*}

To proceed, we consider networks that minimize an energy-like quantity
\begin{align}
\label{E1}
 E(\lambda)=\lambda W + (1-\lambda) e,
\end{align}
where $W$ gives the amount of wire used to connect the network, $e$ the eigenratio, and the parameter $\lambda$ measures the relative weight of the cost of wire and desirablity of superior synchronization properties in the cost function. The limiting cases $\lambda=0$ correspond to no cost for wire in which case a fully connected network with $e=1$ realizes the optimum and $\lambda=1$ in which a linear chain is optimal. Below we use the formalism of Eq. (\ref{E1}) as a convenient description to investigate networks with optimal synchronization properties for differing pressure from spatial constraints. We have, however, also explored optimizing networks for synchronization for given amounts of wire, a procedure that essentially corroborates the results presented below.

In the following we construct optimal network configurations with a numerical optimization scheme via simulated annealing. In principle, the scheme is similar to schemes employed in other studies like \cite{Donetti,Colizza,MB0} which have investigated optimal network topologies in other contexts. In its essentials, the scheme works as follows: an alteration in the network arrangement, being either the addition or deletion of a link or the rewiring of a link to a link vacancy is suggested. Then the `fitness' $E_\text{r}(\lambda)$ of the new network is calculated and the altered configuration is accepted with probability $q=1$ if $E_\text{r}(\lambda)<E(\lambda)$ or with probability $q\propto \exp(\beta (E(\lambda)-E_\text{r}(\lambda)))$ otherwise. The parameter $\beta$, which is gradually increased during the optimization, gives the inverse temperature of the annealing procedure. We repeat the guided rewiring mechanism till no improvement in the network's fitness was found for a number of $T=10 L$ configurations. It is important to note that the number of links, the network organization and the distribution of link lengths are all emergent properties from the optimization.

Numerically, the cost of the optimization increases steeply with the number of simulated oscillators $N$. Further, spatial effects only become apparent if the average spatial distances are not too small compared to the system size. Since on a $d$-dimensionsional grid with periodic boundary conditions the maximum spatial distance scales as $d_\text{max}=\delta \sqrt{d}/2 N^{1/d}$, requirements for the system size grow steeply with $d$. For this reason we limit the paper to $d=1$ where link lengths up to $\delta N/2$ can be observed. As the networks analyzed below have been constructed by a numerical optimization procedure, it is not possible to prove that they represent global minima of the energy defined in Eq. (\ref{E1}). However, we repeated the stochastic optimization with different initial networks and found that the optimized configurations were always structurally similar. 
\begin{figure}
\begin{center}
\includegraphics [width=.45\textwidth]{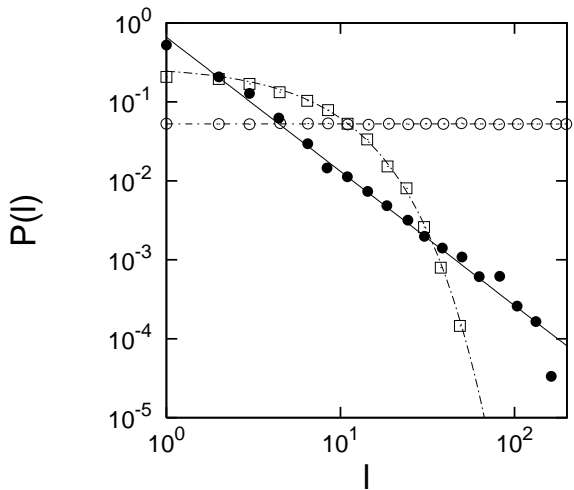}
\caption{Distribution of link length for optimal networks with $N=400$ constructed for $\lambda=0.9$ (filled circles). The data represent averages over 10 networks. The dotted line indicates an inverse power law with exponent $\alpha=1.7$, cf. the solid line. For comparison, also the link length distribution for random networks with the same degree distributions and no constraint on wire (open circles) and for random networks with the same degree distribution and same total length of wire (open boxes) are given. As expected, the link length distribution is uniform for the first and exponential for the second, see dotted lines.}
\label{F1}
\end{center}
\end{figure}

\begin{figure} [tbp]
\begin{center}
\includegraphics [width=.45\textwidth]{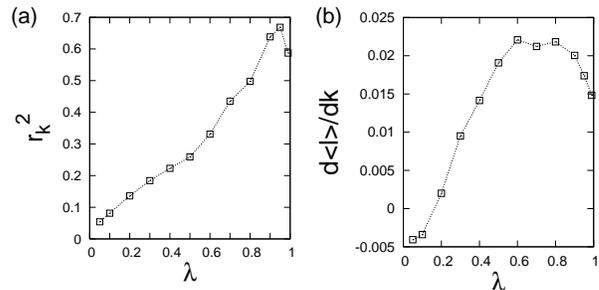}
\caption{Dependence of (a) the ratio $r_\text{k}^2$ between the variances of the degree distributions of the optimal and that of random networks with the same number of links and the same amount of wire and (b) the average pairwise slope of the average link length-degree dependence on $\lambda$. Data for $N=100$ averaged over 100 constructed networks.}
\label{F4}
\end{center}
\end{figure}

By giving a series of optimal example network configurations for different $\lambda$ and the link length distributions for the respective ensembles of optimal networks, Figure \ref{F3}(a)-(h) illustrates some details of the simulation results. For small $\lambda=.05$, cf. panels (a) and (e), many connections are formed. Most links are small, however also longer links are found and the link length distribution is approximately exponential. When $\lambda$ is increased, the increased weight of the cost of wire in the cost function Eq. (\ref{E1})  allows for less links to be formed. By the same token, the link length distribution becomes more skewed. Remarkably, in the range of $0.1\leq\lambda\leq .9$ the link length distributions fit inverse power laws $P(l)\sim l^{-\alpha}$, cf. panels (f) and (g). As expected, costlier wire leads to an increasing dominance of short links and thus larger exponents $\alpha$. When wire becomes very expensive, cf. $\alpha=0.95$ in panels (d) and (h), constraints from the requirement for networks to be connected manifest themselves. The network to be connected requires additional small links, such that deviations from the power law are found for small link lengths, which are excessively abundant.

To substantiate the important finding that optimal networks where spatial constraints are relevant are SWs with a link length distribution that follows a power law, we have also constructed some larger networks with $N=400$. Figure \ref{F1} gives the link length distribution for these networks, clearly indicating a power law tail that holds over several orders of magnitude. For comparison, the figure also shows the link length distributions for two ensembles of reference networks: (i) random networks with the same number of links and (ii) random networks with the same number of links and the same amount of wire. Both ensembles are constructed from rewiring the optimal network configurations. As one expects, the link length distribution becomes uniform in the case of the first ensemble and exponential for the second ensemble. Ensemble (ii) is also a reference ensemble of random networks to evaluate the structure of the optimal configurations.

We continue with an analysis of the structure of the optimal networks, concentrating on the range $.1<\lambda<.9$ for which neither the requirement for synchronization nor the demand to minimize the amount of wire dominate. Comparing to random networks with the same number of links and the same amount of wire, the optimal networks are found to be smaller, have substantially lower clustering coefficients, and strongly disassortative. All these statistics of network organization have been associated with superior synchronization before \cite{synrev}. Additionally, however, in-signal homogeneity has been indentified as an essential determinant of an enhanced synchronizability of networks \cite{Donetti,Motter}. Importantly, the data in Fig. \ref {F4}a highlight that the spatial constraints prevent the optimal networks from becoming homogeneous. For this purpose we measured the ratio between the degree variances $\sigma^2_k=\sum_i (k_i-\langle k\rangle)^2$ between the optimal and random networks. Even though the ratio is always smaller than one, the optimal networks only become homogenous when spatial constraints cease to be relevant. 

In fact, this residual degree heterogeneity appears to be required for an enhanced synchronizability in space. More detailed analysis shows that the average length $\langle l(k)\rangle$ of a link connected to a node of degree $k$ grows strongly with $k$. The correlation is manifested by a Pearson correlation coefficient between $\langle l(k)\rangle$ and $k$ which gives values in the range between $.5$ and $.7$, showing a decline of the strength of the correlation with decreasing $\lambda$. We further calculated a weighted average pairwise slope $d\langle l(k)\rangle/dk=\sum_{i<j} n_i n_j 1/(j-i)(\langle l(j)\rangle-\langle l(i)\rangle)/\sum_{i<j} n_i n_j$, where $n_i$ gives the number of nodes with degree $i$ and $\langle l(i)\rangle$ is the average over all link length of links incident to nodes with degree $i$. The data, cf. Fig. \ref{F4}b emphasize what we observed for the degree-link length correlation coefficient before: Apart from finite size effects for very large $\lambda$ the stronger the role spatial constraints play in the network optimization, the larger the slope of the $\langle l(k)\rangle$-dependence. Hence, optimal synchronization in spatially constrained networks is achieved via the formation of (relative) hubs, around which long distance connections are centred. 

It is noteworthy that the organization of the hubs is different from what has been observed for distance minimizing small worlds in \cite{Gopal}, in so far as no hierarchical core of hub nodes is formed. In contrast, the hub nodes source information from nodes of average degree at far away locations and distribute it to nodes of small degrees close by.

In summary, in this Rapid Communication we have constructed networks that optimize synchronization for varying spatial constraints. We have shown that such networks are classified by link length distributions with power law tails $P(l)\propto l^{-\alpha}$, with a continuous increase of the steepness $\alpha$ when spatial constraints grow in importance. In constrained networks, optimal synchronization is achieved by a residiual heterogeneity in the degree distribution: `hubs' with more than average connections serving as centres for long range communication.  As an aside, we note that a random walk on a synchrony-optimized SW with power law tail of the link length distribution describes a Levy flight on the underlying space. This is of interest, since Levy flights in space have, for instance, been associated with optimal search strategies \cite{Vis}  or human travel patterns \cite{Geisel}.

This research was undertaken on the NCI National Facility in Canberra,
Australia, which is supported by the Australian Commonwealth Government.

\end{document}